\documentclass[11pt]{article}
\usepackage[english]{babel}
\usepackage{fullpage}
\usepackage{epsfig} 
\usepackage{amssymb}
\usepackage{amsmath}
\usepackage{amsthm}
\usepackage[algo2e,algoruled,vlined]{algorithm2e}
\usepackage{times}
\usepackage{graphicx}

\theoremstyle{plain}
  \newtheorem{theorem}{Theorem}
  \newtheorem{lemma}[theorem]{Lemma}
  
\theoremstyle{definition}
  \newtheorem{definition}[theorem]{Definition}

\theoremstyle{remark}

  \newtheorem{example}{Example}

\def\A{\mathcal{A}}
\def\B{\mathcal{B}}
\def\O{\mathcal{O}}
\def\P{\mathcal{P}}
\def\q#1{\mathtt{#1}}
\def\dq#1{\mathtt{#1}}
\def\h{\q{h}}
\def\1{\q{1}}
\def\0{\q{0}}
\def\m{\q{\#}}
\def\j{\q{\_}}
\def\t{\q{@}}
\newcommand{\llabel}{\mathit{label}}

\def\tbbefore{-0mm}
\def\tbinside{-0mm}
\def\tbafter{-4mm}

\title{A unifying framework for seed sensitivity and its application
  to subset seeds}

\author{Gregory Kucherov
\thanks{INRIA/LORIA,
615, rue du Jardin Botanique, B.P. 101, 54602 
Villers-l\`es-Nancy, France, {\tt Gregory.Kucherov@loria.fr}} 
\and 
Laurent No{\'e}
\thanks{UHP/LORIA, 
615, rue du Jardin Botanique, B.P. 101, 54602 
Villers-l\`es-Nancy, France, {\tt Laurent.Noe@loria.fr}} 
\and 
Mikhail Roytberg\thanks{part of this work has been done during
  a visit to LORIA/INRIA in summer 2004}
\thanks{Institute of Mathematical Problems in Biology,
Pushchino, Moscow Region, Russia, {\tt roytberg@impb.psn.ru}}
}

\date{}

\begin{document}
  \maketitle
  %\twocolumn
  \begin{abstract}
We propose a general approach to compute the seed sensitivity, that can
be applied to different definitions of seeds. It treats
separately three components of the seed sensitivity
problem --  a set of target alignments, an associated probability
distribution, and a seed model -- that are specified by
distinct finite automata. The approach is then applied to a new concept of
{\em subset seeds} for which we propose an efficient automaton
construction. Experimental results confirm that sensitive subset seeds
can be efficiently designed using our approach, and can then be used
in similarity search producing better results than ordinary spaced
seeds. 
  \end{abstract}
  \section{Introduction}
%In the framework of heuristic DNA local alignment procedures,
%seeding methods greatly depends on the seed model as well as the seed
%pattern chosen.

In the framework of pattern
matching and similarity search in biological sequences, seeds specify
a class of short sequence motif which, if shared by two sequences, are
assumed to witness a potential similarity. 
Spaced seeds have been introduced several years ago
\cite{BurkhardtKarkkainen03,PatternHunter02} and have been shown to
improve significantly the efficiency of the search. One of the key
problems associated with spaced seeds is a 
precise estimation of the sensitivity of the associated search method.
This is important for comparing seeds and for choosing most
appropriate seeds for a sequence comparison problem to solve.

The problem of seed sensitivity depends on several
components. First, it depends on the {\em seed model} specifying the
class of allowed seeds and the way that seeds match ({\em hit}) potential
alignments. In the basic case, seeds are specified by
binary words of certain length ({\em span}), possibly with a constraint on
the number of 1's ({\em weight}). However, different extensions of this
basic seed model have been proposed in the literature, such as
multi-seed (or multi-hit) strategies
\cite{GBLAST97,BLAT02,PatternHunter02}, seed families
\cite{PatternHunter04,BuhlerRECOMB04,YangWangChenEtAlBIBE04,KucherovNoeRoytbergTCBB05,XuBrownLiMaCPM04,BrownTCBB05},
seeds over non-binary alphabets
\cite{ChenSungGI03,NoeKucherovBMC04}, vector seeds
\cite{BrejovaBrownVinarWABI03,BrownTCBB05}.

The second parameter is the class of {\em target alignments} that are
alignment fragments that one
aims to detect. Usually, 
these are {\em gapless} alignments of a given length. 
%only {\em gapless} alignments are considered.
%This is justified by the fact that biological alignments contain
%gapless subalignments that are targeted by similarity search algorithms to
%detect the whole alignment. 
Gapless alignments are easy to model, in
the simplest case they are represented by binary sequences in the
match/mismatch alphabet. This representation has been adopted by many
authors
\cite{PatternHunter02,KeichLiMaTromp02,BrejovaBrownVinarJBCB04,ChoiZhang04,BuhlerKeichSunRECOMB03,CZZ04}.
The binary representation, however, cannot distinguish between different types
of matches and mismatches, and is clearly insufficient in the case
of protein sequences. In \cite{BrejovaBrownVinarWABI03,BrownTCBB05},
an alignment is represented by a sequence of real numbers that are
{\em scores} of matches or mismatches at corresponding positions. A
related, but yet different approach is suggested in
\cite{NoeKucherovBMC04}, where DNA alignments are represented by
sequences on the ternary alphabet of
match/transition/transversion. Finally, another generalization of
simple binary sequences was considered in
\cite{KucherovNoePontyBIBE04}, where alignments are required to be
{\em homogeneous}, i.e. to contain no sub-alignment with a score larger
than the entire alignment. 
%This amounts to reduce the set of {\em all}
%binary sequences to those sequences that correspond to local
%alignments of biological sequences found by the commonly used
%programs.

The third necessary ingredient for seed sensitivity estimation is the
probability distribution on the set of target alignments. Again,
in the simplest case, alignment sequences are assumed to obey a
Bernoulli model \cite{PatternHunter02,ChoiZhang04}. In more general
settings, Markov or Hidden Markov models are considered
\cite{BuhlerKeichSunRECOMB03,BrejovaBrownVinarJBCB04}. A different way of defining
probabilities on binary alignments has been taken in
\cite{KucherovNoePontyBIBE04}: all homogeneous alignments of a given
length are considered equiprobable.

Several algorithms for computing the seed sensitivity for different
frameworks have been proposed in the above-mentioned
papers. All of them, however, use a common dynamic programming (DP)
approach, first brought up in \cite{KeichLiMaTromp02}.

In the present paper, we propose a general approach to computing the seed
sensitivity. This approach subsumes the cases
considered in the above-mentioned papers, and allows to deal with new
combinations of the three seed sensitivity parameters. 
%For example, we
%will show how to compute the seed sensitivity combining the
%homogeneity requirement and the Hidden Markov Model distribution.
%
The underlying idea of our approach is to specify each of the three
components -- the seed, the set of target alignments, and the
probability distribution -- by a separate finite automaton.

A deterministic finite automaton (DFA) that recognizes all
alignments matched by given seeds was already used in
\cite{BuhlerKeichSunRECOMB03} for the case of ordinary spaced seeds. 
%to develop an algorithm for computing seed sensitivity. 
%The inter-relationship between automata and seed sensitivity was
%first pointed out by Buhler et
%al.~\cite{BuhlerKeichSunRECOMB03}. The authors have considered a deterministic finite
%automaton (DFA), which recognizes the set of all alignments detected by a
%given spaced seed $\pi$, and used the DFA to compute the seed sensitivity.
In this paper, we assume that the set of target alignments is also
specified by a DFA and, more importantly, that the probabilistic model
is specified by a {\em probability transducer} -- a probability-generating
finite automaton equivalent to HMM with respect to the class of generated
probability distributions.
%In our turn, we assume that the set of target alignments is a finite
%prefix-free language over an alignment alphabet, the set obviously can
%be specified by a finite automaton. Furthermore, similar to \cite{BuhlerKeichSunRECOMB03}, we assume that
%the seed component is such that the set of alignments detected by the
%seed is a regular language specified by a DFA.
%Finally, we assume that the probability distribution is
%specified by a {\em probability transducer} -- a probability-generating
%automaton equivalent to HMM with respect to the class of generated
%probability distributions.

%to be solved for any model is a
%computation of seed sensitivity, i.e. the probability, that a given seed
%detects a random alignment, for a target set of alignments, and
%a probability distribution on this set.  The algorithms solving
%this problem  for different models have many details  in common,
%e.g. all of them are DP algorithms. However, they were presented by
%authors as almost independent ones. Our aim is to reveal a common core
%of these algorithms and to propose a new algorithm(s), solving the seed
%sensitivity problem for the class of homogeneous target
%alignments.
%{What about vector seeds? E.g. seed detects a similarity
%  if  weighted sum is >= cut-off}

We show that once these three automata are set, the seed sensitivity
can be computed by a unique general algorithm. This algorithm reduces
the problem to a computation of the total weight over all paths in an
acyclic graph corresponding to the automaton resulting from the
product of the three automata. This computation
can be done by a well-known dynamic
programming algorithm \cite{AhoHopcroftUllman74,FinkRoytberg93} with the time
complexity proportional to the number of transitions of the
resulting automaton. Interestingly, all
above-mentioned seed sensitivity algorithms considered by different
authors can be reformulated as instances of this general algorithm.

%We propose a general framework that allows to reduce the seed
%sensitivity problem to a problem of performing a certain computation
%on a graph. Let $D$ be the set of all target alignments detectable by
%a given seed $S$, for all known models the set $D$ is regular. We
%reduce the seed sensitivity problem to the computation of partition
%function-like sum [standard term?] for the graph of the finite
%automaton $A$ accepting $D$. The latter problem can be solved by the
%well-known DP algorithm, its time and space complexity proportional to
%the number of transitions of the automaton (??). In all known models the automaton
%$A$ can be constructed as a Cartesian product of two automata, $B$ and
%$C$. The automaton $B$ accepts the set of all target alignments. In
%the same time, weights assigned to edges of the automaton $B$
%determine probabilities of the alignments. The automaton $C$ accepts
%the set off all (not necessarily target) alignments, that can be
%detected by the seed $S$. The decomposition leads to the estimation of
%complexity of the automaton $A$.  Therefore, it also allows one to
%estimate the complexity of the algorithm solving the corresponding
%seed sensitivity problem. Based on this approach, we have designed an
%algorithm calculating the seed sensitivity with respect to the set of
%homogeneous alignments for the case of Markov probability model.

In the second part of this work, we study a new concept of 
{\em subset seeds} -- an extension of spaced seeds that allows to deal
with a non-binary alignment alphabet and, on the other hand, still
allows an efficient hashing method to locate seeds. For this definition
of seeds, we define a DFA with a number of states independent of the
size of the alignment alphabet. Reduced to the case of ordinary spaced
seeds, this DFA construction gives the same worst-case number of
states as the Aho-Corasick DFA used in
\cite{BuhlerKeichSunRECOMB03}. Moreover, our DFA has always no
more states than the DFA of \cite{BuhlerKeichSunRECOMB03}, and
has substantially less states on average. 

Together with the general approach proposed in the first part, our DFA
gives an efficient algorithm for computing the sensitivity of subset
seeds, for different classes of target alignments and different
probability transducers. In the experimental part of this work, we confirm
this by running an implementation of our algorithm in order to design
efficient subset seeds for different probabilistic models, trained on
real genomic data. We also show experimentally that designed subset seeds
allow to find more significant alignments than ordinary spaced seeds
of equivalent selectivity. 

%The paper is organized as follows. We start (section
%\ref{section:Framework}) with the description of our
%approach, and . In section \ref{section:SubsetSeeds}, we consider a new
%seed model, the {\em subset seeds}, and show how to compute seed
%sensitivity for this model.

%\cite{YangWangChenEtAlBIBE04}
%\cite{ChenSungGI03}
%\cite{KucherovNoeRoytbergTCBB05}
%\cite{XuBrownLiMaCPM04}
%\cite{BrownTCBB05}
%\cite{BuhlerRECOMB04}

  %
  \section{General Framework}
\label{section:Framework}

Estimating the seed sensitivity amounts to compute the probability for
a random word (target alignment), drawn according to a given
probabilistic model, to belong to a given language, namely the
language of all alignments matched by a given seed (or a set of seeds).

%[???] In this section, we give an automata formalization of each of 
%%give some definitions and facts that are necessary to define 
%underlying languages and probability distributions.  
%In particular, we define a general way to specify probabilities
%on words, based on probability transducers (section
%\ref{subsection:proba-assign}). Together with a DFA defining the set
%of alignments detected by a given seed model (section
%\ref{subsection:seed-autom}), we introduce in section
%\ref{subsection:algo} a general method to compute the seed
%sensitivity. 
% allows one to use a DP
%algorithm to compute probabilities of finite languages (section
%\ref{subsection:seed-autom}). Finally, we give a detailed
%description of our approach (section \ref{subsection:algo}).

\subsection{Target Alignments}

Target alignments are represented by words over an alignment alphabet
$\A$. In the simplest case, considered most often, the alphabet is
binary and expresses a match or a mismatch occurring at each alignment
column. However, it could be useful to consider larger alphabets,
such as the ternary alphabet of match/transition/transversion for the
case of DNA (see 
\cite{NoeKucherovBMC04}). The importance of this extension is even
more evident for the protein case (\cite{BrownTCBB05}), where
different types of amino acid pairs are generally distinguished. 

Usually, the set of target alignments is a finite set. In the case
considered most often
\cite{PatternHunter02,KeichLiMaTromp02,BrejovaBrownVinarJBCB04,ChoiZhang04,BuhlerKeichSunRECOMB03,CZZ04},
target alignments are all words of a given length $n$. This set is 
trivially a regular language that can be specified by a deterministic automaton
with $(n+1)$ states. % (probabilistic alignment model ??)
%On the other hand, the probability $Prob(w)$ of a word $w \in A^n$ is
%specified by an appropriate weighted automaton (corresponding e.g. to
%a Bernoulli or Markov model). 
%
%Target alignments are represented as words on a given alphabet $\A$ 
However, more complex definitions of target alignments have been
considered (see e.g. \cite{KucherovNoePontyBIBE04}) that aim to capture more
adequately properties of biologically relevant alignments.
%The language $L_T$ is the set of all target alignments.  
In general, we assume that the set of target alignments is a finite regular
language $L_T\in\A^*$ 
and thus can be represented by an acyclic DFA 
$T = <Q_T,q_T^0,q_T^F,\A,\psi_T>$.

%Homogeneity principle  only considers alignments of score over a given
%threeshold: this implies to provide a scoring function ($\A \to \Z$). 
%Moreover, the Homogeneity restricts the alignment set to MSP
%alignments (it means that no sub-alignment of any alignment can give a
%larger score than the complete alignment).

\subsection{Probability Assignment}
\label{subsection:proba-assign}

Once an alignment language $L_T$ has been set, we have to define a
probability distribution on the words of $L_T$. We do this 
using probability transducers.

%\subsubsection{Probability transducer}
%\label{subssubection:probabilitytransducer}

A probability transducer is a finite automaton without final states in
which each transition outputs a {\em probability}. 
\begin{definition}
  \label{definition:probabilitytransducer}
  A {\em probability transducer} $G$ over an alphabet $\A$ is a 4-tuple
  $<Q_G,q_G^0,\A, \rho_G>$, where $Q_G$ is a finite set of states, $q_G^0\in Q_G$ is
  an initial state, and $\rho_G: Q_G\times \A\times Q_G\rightarrow [0,1]$ is
  a real-valued probability function such that \newline
  \mbox{$\forall q\in Q_G, \sum_{q'\in Q_G, a \in \A} \rho_G(q,a,q') = 1$}.
\end{definition}
A {\em transition} of $G$ is a triplet $e=<q,a,q'>$ 
%($q,q' \in Q_G, a \in \A$)
such that $\rho(q,a,q')>0$. Letter $a$ is called the {\em label} of
$e$ and denoted $\llabel(e)$. A probability transducer $G$ is {\em deterministic}
if for each $q\in Q_G$ and each $a\in\A$, there is at most one
transition $<q,a,q'>$. 
%If $e=<q, a, q'>$ is a transition, then $q\in Q_G$ is called its
%{\em source state}, $q'\in Q_G$ its {\em destination state}, and $a \in \A$ its
%{\em label} (denoted $\llabel(e)$).
%A {\em path} $P = \{ e_1, ..., e_n \}$ in $G$ is a sequence of
%transitions, and 
For each path $P = ( e_1, ..., e_n )$ in $G$, we define its 
{\em label} to be the word
$\llabel(P) = \llabel(e_1)... \llabel(e_n)$, and the associated
probability to be the product $\rho(P) = \prod_{i=1}^n \rho_G(e_i)$.
A path is {\em initial}, if its start state is the initial state
$q_G^0$ of the transducer $G$.

\begin{definition}
  \label{definition:weightword}
The {\em probability} of a word $w\in \A^*$ according to a probability
transducer $G = <Q_G, q^0_G, \A,\rho_G>$, denoted $\P_G(w)$, is the
sum of probabilities of all initial paths in $G$ with the label
$w$. $\P_G(w)=0$ if no such path exists. 
%Let $L\subseteq \A^*$ be a finite language. 
The probability $\P_G(L)$ of a finite language $L\subseteq \A^*$ 
according a probability transducer $G$ is defined by $\P_G(L) =
\sum_{w\in L} \P_G(w)$.
\end{definition}

%[???] Informally, we represent the probability of a word $w$ as the sum of
%complementary events corresponding to different initial paths $P$ with
%$label(P) = w$.
%%%%% do we need this sentence ?? 
Note that for any $n$ and for $L=A^n$ (all words of length $n$), $\P_G(L) =1$. 

%\subsubsection{Examples of Probability Transducers}

Probability transducers can express common probability distributions on
words (alignments). 
Bernoulli sequences with independent probabilities of each symbol
\cite{PatternHunter02,ChoiZhang04,CZZ04} 
can be specified with deterministic one-state probability transducers. 
%In first-order
%Markov sequences, the probability of each symbol depends only on ...
%Two commonly used generalizations of Markov sequences are Markov
%sequences of order $k$  ($MS[k]$ model) and hidden Markov models (HMM)\cite{Rabiner89}.
In Markov sequences of order $k$
\cite{BuhlerKeichSunRECOMB03,BuhlerRECOMB04}, the probability of each
symbol depends on $k$ previous symbols. They can therefore be
specified by a deterministic probability transducer with at most
$|\A|^{k}$ states.

A Hidden Markov model (HMM) \cite{BrejovaBrownVinarJBCB04}
corresponds, in general, to a non-deterministic probability
transducer. The states of this transducer correspond to the (hidden)
states of the HMM, plus possibly an additional initial state. Inversely, for
each probability transducer, one can construct an HMM generating the
same probability distribution on words. Therefore, non-deterministic
probability transducers and HMMs are equivalent with respect to the
class of generated probability distributions. The proofs are
straightforward and are omitted due to space limitations. 
%Other probability transducers as M3, M8 have also been proposed in ~\cite{???}
%to mimic alignment phase regularities in DNA coding sequences.

\subsection{Seed automata and seed sensitivity}
\label{subsection:seed-autom}

Since the advent of spaced seeds
\cite{BurkhardtKarkkainen03,PatternHunter02}, different extensions of
this idea have been proposed in the literature (see Introduction). For
all of them, the set of possible alignment fragments matched by a seed
(or by a set of seeds) is a finite set, and therefore the set of matched
alignments is a regular language. 
For the original spaced seed model, this observation was used by Buhler
et al.~\cite{BuhlerKeichSunRECOMB03} who proposed an algorithm for
computing the seed sensitivity based on a DFA defining the language of
alignments matched by the seed. 
In this paper, we extend this approach to a general one that allows a
uniform computation of seed sensitivity for a wide class of settings
including different probability distributions on target alignments, as
well as different seed definitions. 

Consider a seed (or a set of seeds) $\pi$ under a given seed model. 
We assume that the set of alignments $L_\pi$ matched 
%(or {\em detected}, or {\em hit}) 
by $\pi$
is a regular language recognized by a DFA
$S_\pi=<Q_S,q_S^0,Q_S^F,\A,\psi_S>$. Consider a finite set $L_T$ of
target alignments and a probability transducer $G$. 
Under this assumptions, the
sensitivity of $\pi$ is defined as the conditional probability 
\begin{equation}
\label{cond-proba}
\frac{\P_G(L_T \cap L_\pi)}{\P_G(L_T)}. 
\end{equation}
%of the language $L = L_T \cap L_\pi$ according to the
%probability distribution specified by a probability transducer $G$. 
%
%$L$ is the language of all target alignments detectable by a given
%seed $\pi$. 
%$L_T$ is the set of all target alignments and $L_\pi$ is the set of all
%alignments detectable by $\pi$. 
%Usually, both $L_T$ and $L_\pi$ are regular languages.

An automaton recognizing $L=L_T \cap L_\pi$ can be obtained 
as the product of automata $T$ and $S_\pi$ recognizing $L_T$ and
$L_\pi$ respectively. Let $K = <Q_K,q_K^0,Q_K^F,\A,\psi_K>$ be this
automaton. 
We now consider the product $W$ of 
$K$ and $G$, denoted $K \times G$, defined as follows. 

\begin{definition}
Given a DFA 
$K = <Q_K,q_K^0,Q_K^F,\A,\psi_K>$ and a probability transducer $G=<Q_G, q^0_G, \A,\rho_G>$,
the product of $K$ and $G$ is the {\em probability-weighted automaton}
 $W =<Q_W,q_W^0,Q_W^F,\A,\rho_W>$ (for short, {\em PW-automaton}) such that 
%
% Clarifier les notions 'automaton', 'probability transducer', 'weighted automaton'
%
\begin{itemize}
\item $Q_W   =  Q_K \times Q_G$,
\item $q_W^0 =  (q_K^0, q_G^0)$,
\item $q_W^F =  \{ (q_K,q_G) | q_K \in Q_K^F\}$,
\item $\rho_W((q_K, q_G),a,(q'_K,q'_G))  = \begin{cases} \rho_G(q_G,
    a, q'_G) & 
  \mbox{ if } \psi_K(q_K,a) = q'_K, \\ 
%\mbox{ and } \rho_W((q_K,q_G),a,(q'_K,q'_G))=
0 & \mbox{ otherwise.}
\end{cases}$
%$\{<\!(q_K, q_G),a,(q'_K,q'_G)\!>  |\:  \psi_K(q_K, a) =
%  q'_K \mbox{ and } \rho_G(q_G, a, q'_G)>0 \}$, in this case
%  $\varphi_W = \rho_G(q_G, a, q'_G)$,
%%%%%%%% CHECK OUT this definition. 
\end{itemize}
\end{definition}
$W$ can be viewed as a non-deterministic probability transducer with
final states. $\rho_W((q_K,q_G),a,(q'_K,q'_G))$ is the 
{\em probability} of the transition $<(q_K,q_G),a,(q'_K,q'_G)>$. 
A path in $W$ is called {\em full} if it goes from the initial to a final
state. % The following lemma 
%%%%%%%% Define how the automaton works (path, label of a path,
%%%%%%%% acceptance) ???
\begin{lemma}
\label{lemma:LangWeight}
Let $G$ be a probability transducer.
Let $L$ be a finite language and $K$ be a deterministic automaton
recognizing $L$. Let $W = G \times K$. 
The probability $\P_G(L)$ 
%of language $L$ according to the probability transducer $G$ 
is equal to sum of probabilities of all full paths in $W$. 
\end{lemma}
\begin{proof}
Since $K$ is a deterministic automaton, each word $w\in L$ corresponds
to a single accepting path in $K$ and the paths in $G$ labeled $w$ (see
Definition~\ref{definition:probabilitytransducer}) are in one-to-one
correspondence with the full path in $W$ accepting $w$. 
By definition, $\P_G(w)$ is equal to the
sum of probabilities of all paths in $G$ labeled $w$. Each such path
corresponds to a unique path in $W$, with the same probability. Therefore,
the probability of $w$ is the sum of probabilities of corresponding paths in
$W$. Each such path is a full path, and paths for distinct words $w$
are disjoint. The lemma follows.
\end{proof}
%%%%%%%% ligthen the proof ???

\subsection{Computing Seed Sensitivity}
\label{subsection:algo}

Lemma \ref{lemma:LangWeight} reduces the computation of seed
sensitivity to a computation of the sum of probabilities of paths in a
PW-automaton. 

%L1> NORMALISE R R_\pi notations
\begin{lemma}
\label{lemma:DP}
Consider an alignment alphabet $\A$, a finite set $L_T \subseteq \A^*$
of target alignments, and a set $L_{\pi} \subseteq \A^*$ of all
alignments matched by a given seed $\pi$. 
Let $K = <Q_K, q_t^0, Q_K^F, \A, \psi_Q>$ 
be an acyclic DFA recognizing the language $L=L_T \cap L_\pi$. Let
further $G = <Q_G, q_G^0, \A, \rho >$ be a probability transducer
defining a probability distribution on the set $L_T$. Then $\P_G(L)$
can be computed in time  
%Then the sensitivity
%\begin{equation}\label{equation:SeedSensitivity}
%  s_{\pi} = Prob(R_\pi)
%\end{equation}
%can be found by a DP algorithm with time complexity
\begin{equation}\label{equation:SeedSensitivityTimeComplexity}
  \O(|Q_G|^2\cdot|Q_K|\cdot|\A|)
\end{equation}
and space %complexity
\begin{equation}\label{equation:SeedSensitivitySpaceComplexity}
  \O(|Q_G|\cdot|Q_K|).
\end{equation}
\end{lemma}

\begin{proof} By Lemma~\ref{lemma:LangWeight}, the probability
%  (\ref{equation:SeedSensitivity}) 
of $L$ with respect to $G$
can be computed as the sum of probabilities of all full paths in $W$. Since
$K$ is an acyclic automaton, so is $W$. Therefore, the sum of probabilities
of all full paths in $W$ leading to final states $q^F_W$ can  be
computed by a classical DP algorithm \cite{AhoHopcroftUllman74}
applied to acyclic directed graphs 
(\cite{FinkRoytberg93} presents a survey of application of this
technique to different bioinformatic  problems). 
The time complexity of the algorithm is proportional to the number of 
transitions in $W$. $W$ has $|Q_G|\cdot |Q_K|$ states, and for each
letter of $\A$, each state has at most $|Q_G|$ outgoing
transitions. The bounds follow. 
\end{proof}

Lemma~\ref{lemma:DP} provides a general approach to compute the seed
sensitivity. To apply the approach, one has to define three automata:
\begin{itemize}
\item a deterministic acyclic DFA $T$ specifying a set of
  target alignments over an alphabet $\A$ (e.g. all words of a given
  length, possibly verifying some additional properties), 
\item a (generally non-deterministic) probability transducer $G$ specifying
  a probability distribution on target alignments (e.g. Bernoulli
  model, Markov sequence of order $k$, HMM), 
\item a deterministic DFA $S_\pi$ specifying the seed model via a set
  of matched alignments.
\end{itemize}
As soon as these three automata are defined, Lemma~\ref{lemma:DP} can
be used to compute probabilities $\P_G(L_T\cap L_\pi)$ and $\P_G(L_T)$
in order to estimate the seed sensitivity according to (\ref{cond-proba}).

%The automata $S_\pi$ and $T$ have to be deterministic, their product
%$S_\pi \times T$ has to be acyclic.

Note that if the probability transducer $G$ is deterministic (as it is the
case for Bernoulli models or Markov sequences), then the time
complexity (\ref{equation:SeedSensitivityTimeComplexity}) is
$\O(|Q_G|\cdot|Q_K|\cdot|\A|)$. 
In general, the
complexity of the algorithm can be improved by reducing the involved
automata. Buhler et al. \cite{BuhlerKeichSunRECOMB03} introduced the
idea of using the Aho-Corasick automaton~\cite{AhoCorasick74} as the seed automaton
$S_\pi$ for a spaced seed. The authors
of \cite{BuhlerKeichSunRECOMB03} considered all binary alignments of a
fixed length $n$ distributed according to a Markov model of order $k$. In
this setting, the obtained complexity was $\O(w2^{s-w}2^k n)$, where
$s$ and $w$ are seed's span and weight respectively. Given that the
size of the Aho-Corasick automaton is $\O(w2^{s-w})$, this complexity
is automatically implied by
Lemma~\ref{lemma:DP}, as the size of the
probability transducer is $\O(2^k)$, and that of the target alignment
automaton is $\O(n)$. Compared to \cite{BuhlerKeichSunRECOMB03},
our approach explicitly distinguishes the descriptions of matched
alignments and their probabilities, which allows us to automatically
extend the algorithm to more general cases. 

Note that the idea of using the Aho-Corasick automaton can be applied
to more general seed models than individual spaced seeds (e.g. to
multiple spaced seeds, as pointed out in
\cite{BuhlerKeichSunRECOMB03}). In fact, all currently proposed seed
models can be described by a finite set of matched alignment
fragments, for which the Aho-Corasick automaton can be constructed. We
will use this remark in later sections. 

%Note that the complexity of the presented seed sensitivity algorithm
%is governed by numbers of states and transitions in the reduced  form
%of the weighted automaton $W = K \times G$. However, the representation of the
%automaton $W$ used to obtain the bounds
%(\ref{equation:SeedSensitivityTimeComplexity})-(\ref{equation:SeedSensitivitySpaceComplexity})
%and
%(\ref{equation:SeedSensitivityTimeComplexity2})-(\ref{equation:SeedSensitivitySpaceComplexity2})
%in general is not in reduced form. Therefore, in some important special
%cases the bounds can be significantly improved, cf.~\cite{BuhlerKeichSunRECOMB03}.

The sensitivity of a spaced seed with respect to an HMM-specified
probability distribution over
% and target sets consisting of all binary words 
binary target alignments of a given length $n$
was studied by Brejova et al. \cite{BrejovaBrownVinarJBCB04}. 
The DP algorithm of \cite{BrejovaBrownVinarJBCB04} has a lot in common
with the algorithm implied by
Lemma~\ref{lemma:DP}. In 
particular, the states of the algorithm of
\cite{BrejovaBrownVinarJBCB04} are triples 
$<w, q, m>$, where $w$ is a prefix of the seed $\pi$, $q$ is a state
of the HMM, and $m\in [0..n]$.  
The states therefore correspond to the construction implied by
Lemma~\ref{lemma:DP}. 
%The states are therefore in correspondence with the states of weighted
%automaton for the model under consideration i.e.
%the state  $<w, q, m>$ of the algorithm of Brejova et al corresponds
%to the state $<w, q, m+|w|>$ of the algorithm given in section
%~\ref{subsection:TheMainAlgorithm}.
%The recursive equations of the algorithm of Brejova et al
%and the version of the automaton-based
%algorithm are also similar, and the main terms of run-time
%bounds of the algorithms are the same. 
However, the authors
of~\cite{BrejovaBrownVinarJBCB04} do not consider any automata,
%and do not use the Aho-Corasic algorithm for the preprocessing step. 
%Thus, the step is more time consuming than the one of the
%automaton-based algorithm. 
which does not allow to optimize the preprocessing step (counterpart
of the automaton construction) and, on the other hand, does not allow to extend the algorithm to more
general seed models and/or different sets of target alignments. 

%In Brejova et al~\cite{BrejovaBrownVinarWABI03} the algorithm is
%generalized to work out vector seeds. The relationship between the
%algorithm of~\cite{BrejovaBrownVinarWABI03} and the automaton-based
%algorithm is analogous to that of the algorithm
%~\cite{BrejovaBrownVinarJBCB04}.

%L1>
%Until now, we have outlined the framework used for the seed sensitivity
%estimation, explaining how probability models, alignments, and seed, are
%represented in this layout. Usually probabilities are assigned by classical models
%(Markov,HMM) or more specifical models ($M_3$,$M_8$,~\cite{BrejovaBrownVinarJBCB04}).
%Alignments are frequently considered as words of fixed size, set possibly
%restrained by use of constraints (scoring constraints,
%homogeneity constraints ~\cite{KucherovNoePontyBIBE04}).

%The main problem comes from the seed model description. Since one has
%to consider a large set of seeds to find an efficient one (possibly
%the best seed of the set), seeds have to be concisely described, due to
%processing costs involved.
%A good approach to do this is to minimize seed automaton
%~\cite{BuhlerKeichSunRECOMB03}, but since the initial automaton has to
%be built explicitly before minimization, efficient design of it is
%also welcome.

%In the next section, we propose a new seed model fitted to direct
%indexing (Subset seed model), which is an extension of classical
%spaced seeds. We also provide a new initial automaton fitted to this
%model, but also to classical spaced seeds, since it generates for each
%seed an initial automaton with a smaller number of states.
%<L1

A key to an efficient solution of the sensitivity problem remains the
definition of the seed. It should be expressive enough to be able to
take into account properties of biological sequences. On the other
hand, it should be simple enough to be able to locate seeds fast and
to get an efficient algorithm for computing seed
sensitivity. According to the approach presented in this section, the
latter is directly related to the size of a DFA specifying the seed. 

  \section{Subset seeds}
\label{section:SubsetSeeds}

\subsection{Definition}
\label{subsection:SubsetSeeds}

%Presently most popular seed models are spaced
%seeds~\cite{BurkhardtKarkkainen03,PatternHunter02} and vector
%seeds~\cite{BrejovaBrownVinarWABI03}. 

Ordinary spaced seeds use the simplest possible binary ``match-mismatch''
alignment model that allows an efficient implementation by hashing all
occurring combinations of matching positions. 
%The  spaced seed consider alignment positions
%independently, claiming the match on all ``essential''
%positions, and are particularly fitted for direct indexing schemes.
A powerful generalization of spaced seeds, called {\em vector seeds},
has been introduced in \cite{BrejovaBrownVinarWABI03}. 
Vector seeds allow one to use an arbitrary alignment alphabet
%that is of benefit in many cases. Another difference between
%the models is that the vector seed weight method is 
and, on the other hand, provide a flexible definition of a hit based
on a cooperative contribution of seed positions. 
%defines a hit through the integral
%contribution of all seed positions rather than considering
%distinguished positions independently, as spaced seeds do. 
%``integral'',i.e. it weights
%the entire putative match, rather than its individual positions.
A much higher expressiveness of vector seeds lead to more complicated 
algorithms and, in particular, prevents the application of direct
hashing methods at the seed location stage. 
%Formally speaking, spaced seeds are special case of vector seeds. But
%more wide potentialities of vector seeds lead to more complicated
%algorithms. As usual, one has to use not the most powerful model, but
%the model that is the most adequate one for the problem under
%consideration. 

In this section, we consider {\em subset seeds} that have an 
intermediate expressiveness between spaced and vector seeds. It allows
an arbitrary alignment alphabet and, on the other hand, still allows
using a direct hashing for locating seed, which maps each
string to a unique entry of the hash table. 
We also propose a construction of a seed automaton for subset seeds,
different from the Aho-Corasick automaton. The automaton has
$\O(w 2^{s-w})$ states {\em regardless of the size of the alignment
  alphabet}, where $s$ and $w$ are respectively the span of the seed
and the number of ``must-match'' positions. From the general algorithmic
framework presented in the previous section (Lemma~\ref{lemma:DP}),
this implies that the seed sensitivity can be computed for subset
seeds with same complexity as for ordinary spaced seeds. 
Note also that for the binary alignment
alphabet, this bound is the same as the one implied by the Aho-Corasick
automaton. However, for larger alphabets, the Aho-Corasick construction
leads to $\O(w|\A|^{s-w})$ states. In the experimental part of this
paper (section~\ref{subsection:aut-size}) we will show that even for
the binary alphabet, our automaton construction yields a smaller
number of states in practice. 

Consider an alignment alphabet $\A$. We always assume that $\A$
contains a symbol $\1$, interpreted as ``match''. 
A {\em subset seed} is defined as a word over a {\em seed alphabet}
$\B$, such that
\begin{itemize}
\item letters of $\B$ denote subsets of the alignment
alphabet $\A$ containing $\1$ ($\B \subseteq \{\1\}\cup 2^{\A}$), 
\item  $\B$ contains a letter $\m$ that denotes subset $\{\1\}$,
\item a subset seed $b_1 b_2\ldots b_m\in\B^m$ matches an alignment fragment
$a_1 a_2\ldots a_m\in\A^m$ if $\forall i \in [1..m]$, $a_i \in b_i$. 
\end{itemize}

% is a triple $<\B,\A>$, where the
%seed alphabet $\B$, the alignment alphabet $\A$, and the matching
%function $\beta$ meet the following conditions.
%\begin{itemize}
%% --Introduce the similarity/seed alphabet
%\item each letter $b$ of the alphabet $\B$
%represents a non empty subset of $\A$; for the sake of brevity, we
%will write  
%$\B \subseteq 2^{\A}  \setminus \{ \emptyset\}$ ($\B$ consists of
%non-empty subsets of the alphabet $\A$); 
%%L1> la notation  $\B \subseteq 2^{\A}$ ne m'est pas familière du
%%tout. Est elle classique ?
%\item a seed $b_1 b_2\ldots b_n$ matchs an alignment fragment $a_1 a_2\ldots a_n$
%{\bf if $\forall k \in [1,n] \; a_k \in b_k$};
%\item there is  a ``match'' symbol $\1 \in  \A$, that fits all the
%  subsets from $\B$ ( $\forall  b \in \B  (\1 \in b)$ );
%\item there is  a ``must-match'' symbol  $\m \in \B$ representing the subset $\{\1\}$, this is the
%minimal subset in $\B$.
%\end{itemize}

The {\em $\m$-weight} of a subset seed $\pi$ is the number of $\m$
in $\pi$ and 
%In case of spaced seeds, $\m$-weight of a seed is equal to its
%weight. 
the {\em span} of $\pi$ is its length.

% --Transition constrained seed example
%\paragraph{Example 1.} 
\begin{example}
\label{example-subset}
\cite{NoeKucherovBMC04} considered
the alignment alphabet $\A = \{\1,\h,\0\}$ representing respectively a
match, a transition mismatch, or a transversion mismatch in a DNA
sequence alignment. The seed alphabet is $\B = \{\m,\t,\j\}$ denoting
respectively subsets $\{\1\}$, $\{\1,\h\}$, and $\{\1,\h,\0\}$. 
Thus, seed $\pi = \dq{\#@\_\#}$ matches alignment $s = \dq{10h1h1101}$
at positions $4$ and $6$. The span of $\pi$ is $4$, and the
$\m$-weight of $\pi$ is 2. 
\end{example}
Note that unlike the weight of
ordinary spaced seeds, the $\m$-weight cannot serve as a measure of
seed selectivity. In the above example, symbol $\t$ should be assigned
weight $0.5$, so that the weight of $\pi$ is equal to $2.5$
(see~\cite{NoeKucherovBMC04}).

\subsection{Subset Seed Automaton}
% -- Fix our notation for the rest of the description
Let us fix an alignment alphabet $\A$, a seed alphabet $\B$, and a
seed $\pi=\pi_1\pi_2\ldots \pi_{m} \in \B^*$ of span $m$ and $\m$-weight
$w$. Let $R_{\pi}$ be the set of all non-$\m$ positions in
$\pi$, $|R_{\pi}|= r = m - w$.
%
% -- The state definition
We now define an automaton 
$S_{\pi} = <Q, q_0, Q_f, \A,\psi : Q\times\A\to Q>$ 
that recognizes the set of all alignments matched by $\pi$. 

The states $Q$ of $S_{\pi}$ are pairs $<X,t>$ such that 
$X \subseteq R_{\pi}, t \in [ 0,\ldots,m ]$, with the following
invariant condition. 
Suppose that $S_{\pi}$ has read a prefix $s_1\ldots s_p$ of an
alignment $s$ and has come to a state $<X,t>$. 
%
% Then $X$
% enumerates all positions of $\pi$ that can match the last
% non-$\1$ symbol of $s$, and $t$ is the length of $\1$-only suffix of
% $s$. 
Then $t$ is the length of the longest suffix of $s_1\ldots s_p$ of the
form $\1^i$, $i\leq m$, 
%{\em last run} of $\1$ (maximal $\1$-suffix of $s$). 
%
and $X$ contains all positions $x_i\in R_\pi$ %(non-$\m$ positions of $\pi$)
such that prefix $\pi_1 \cdots \pi_{x_i}$ of $\pi$ matches a suffix of
$s_{1}\cdots s_{p-t}$. 
%Fig.~\ref{figure:Statedef} illustrates this
%definition with an example. \\ 
% 
%Therefore, all positions $x \in X$ correspond to non $\m$ positions
%of the seed $\pi$.
% -- Drawing
\def\PiMotif#1#2{
  \put(#1,#2){
     \put(0,0){\large$\mbox{\tt \#}$}
     \put(1,0){\large$\mbox{\tt  @}$}
     \put(2,0){\large$\mbox{\tt \#}$}
     \put(3,0){\large$\mbox{\tt \_}$}
     \put(4,0){\large$\mbox{\tt \#}$}
     \put(5,0){\large$\mbox{\tt \#}$}
     \put(6,0){\large$\mbox{\tt \_}$}
     \put(7,0){\large$\mbox{\tt \#}$}
     \put(8,0){\large$\mbox{\tt \#}$}
     \put(9,0){\large$\mbox{\tt \#}$}
  }
}

\def\StrMotif#1#2{
  \put(#1,#2){
     \put(0,0){\large$\mbox{\tt 1}$}
     \put(1,0){\large$\mbox{\tt 1}$}
     \put(2,0){\large$\mbox{\tt 1}$}
     \put(3,0){\large$\mbox{\tt h}$}
     \put(4,0){\large$\mbox{\tt 1}$}
     \put(5,0){\large$\mbox{\tt 0}$}
     \put(6,0){\large$\mbox{\tt 1}$}
     \put(7,0){\large$\mbox{\tt 1}$}
     \put(8,0){\large$\mbox{\tt h}$}
     \put(9,0){\large$\mbox{\tt 1}$}
     \put(10,0){\large$\mbox{\tt 1}$} 
     \put(11,0){\large$\mbox{\tt ...}$}
  }
}

\def\PisevenPref#1#2{
  \put(#1,#2){
     \put(0,0){\large$\mbox{\tt \#}$}
     \put(1,0){\large$\mbox{\tt  @}$}
     \put(2,0){\large$\mbox{\tt \#}$}
     \put(3,0){\large$\mbox{\tt \_}$}
     \put(4,0){\large$\mbox{\tt \#}$}
     \put(5,0){\large$\mbox{\tt \#}$}
     \put(6,0){\large$\mbox{\tt \_}$}
  }
}

\def\PifourPref#1#2{
  \put(#1,#2){
     \put(0,0){\large$\mbox{\tt \#}$}
     \put(1,0){\large$\mbox{\tt  @}$}
     \put(2,0){\large$\mbox{\tt \#}$}
     \put(3,0){\large$\mbox{\tt \_}$}
  }
}

\def\PitwoPref#1#2{
  \put(#1,#2){
     \put(0,0){\large$\mbox{\tt \#}$}
     \put(1,0){\large$\mbox{\tt  @}$}
  }
}

% -- example to add here
\begin{figure*}[htb]\center
  \begin{picture}(100,40)(-2,-20)\noindent\centering\setlength{\unitlength}{8pt}
    \put(-18, 1){$(a)$}
    \put(-15, 0){\large$\pi = $}\PiMotif{-12}{ 0}% seed motif
    \put(-18,-2){$(b)$}
    \put(-15,-3){\large$ s = $}\StrMotif{-12}{-3}% text 
    \put(10,1){$(c)$}
    \put(22.5, 4){$s_9$}
    \put(23.3, 3.5){\line(0,-1){0.5}}
    % [
    \put(24.0, 3.5){\line(0,-1){0.5}}
    \put(25.6, 3.5){\line(0,-1){0.5}}
    \put(24.0, 3.5){\line(1, 0){1.6}}
    \put(24.7, 4){$t$}
    % ]
    \StrMotif{15}{2}
    \put(13, 0.25){$\pi_{1..7}=$}\PisevenPref{17}{0} %prefixes
    \put(16,-1.25){$\pi_{1..4}=$}\PifourPref{20}{-1.5}
    \put(18,-2.75){$\pi_{1..2}=$}\PitwoPref{22}{-3}
  \end{picture}
\caption{\label{figure:Statedef} Illustration to Example~\ref{example-state}}
\end{figure*}
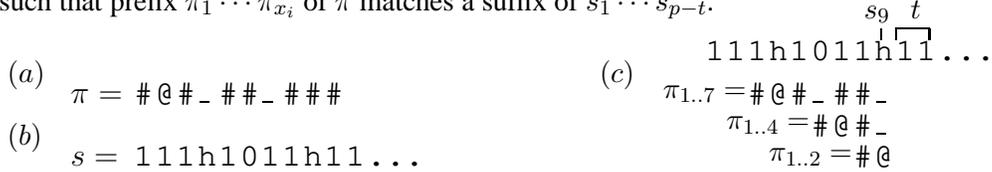

\begin{example}
\label{example-state}
In the framework of Example~\ref{example-subset}, 
consider a seed $\pi$ and an alignment prefix $s$ 
of length $p = 11$ given on Figure~\ref{figure:Statedef}(a) and (b) respectively. The length $t$ of the last run of
$\1$'s of $s$ is $2$. The last mismatch position of $s$ is $s_9 = \h$.
The set $R_\pi$ of non-$\m$ positions of $\pi$ is $\{2,4,7\}$ and 
$\pi$ has 3 prefixes ending at positions of $R_\pi$ (Figure~\ref{figure:Statedef}(c)). 
Prefixes $\pi_{1..2}$ and $\pi_{1..7}$ do match suffixes of 
$s_1 s_2\ldots s_9$, and prefix $\pi_{1..4}$ does not.
Thus, the state of the automaton after reading $s_1 s_2\ldots s_{11}$ is
$<\{2,7\}, 2>$.\\
\end{example}

The initial state $q_0$ of $S_{\pi}$ is the state $ <\emptyset, 0>$.
The final states $Q_f$ of $S_{\pi}$ are all states $q = <X,t>$, where
$max\{X\} + t = m$. All final states are merged into one state.

%Below, we will suppose that all final states are merged into on state.~\\

% -- Transition function $\psi$
The transition function $\psi(q,a)$ is defined as follows: 
If $q$ is a final state, then $\forall a \in \A$, $\psi(q,a) = q$. 
If $q = <X,t>$ is a non-final state, then
\begin{itemize}
\item if $a = \1$ then $\psi(q,a) = <X,t+1>$, 
\item otherwise $\psi(q,a) =  <X_{U} \cup X_{V},0>$ with
  \begin{itemize}
  \item $X_{U} = \{ x           | x \leq t + 1 \mathrm{~and~} a \mathrm{~matches~} \pi_{x} \}$
  \item $X_{V} = \{ x + t + 1   | x \in X  \mathrm{~and~} a \mathrm{~matches~} \pi_{x+t+1}\}$
  \end{itemize}
\end{itemize}

% -- Automaton exactly recognizes the seed language
\begin{lemma}\label{lemma:Gpi}
The automaton $S_{\pi}$ accepts the set of all alignments matched by $\pi$.
\end{lemma}
\begin{proof}
% Straightforward ( Mr Roytberg )
It can be verified by induction that the invariant condition on the states
$<X,t>\in Q$ is preserved by the transition function $\psi$. 
The final states verify $max\{X\} + t = m$, which implies
that $\pi$ matches a suffix of $s_1\ldots s_p$. 
%  of  is equal to $m$ (in this case the seed
%  matches without any mismatch under its jokers), either that the
%  maximal prefix matching covering the last error is completed by a
%  long enough run of $1$, so the seed matches.
\end{proof}

% -- Number of states
\begin{lemma}\label{lemma:GnbStates}
The number of states of the automaton $S_{\pi}$ is no more than $(w+1)2^{r}$.
\end{lemma}
\begin{proof}
Assume that $R_\pi=\{x_1,x_2,\ldots,x_r\}$ % are all non-$\m$ positions of the seed $\pi$
and $x_1 < x_2 \cdots < x_r$. Let $Q_i$ be the set of non-final states
$<X,t>$ with $max\{X\} = x_i$, $i\in [1..r]$.
For states $q = <X,t> \in Q_i$ there are $2^{i-1}$
possible values of $X$ and $m-x_i$ possible values of 
$t$,
%($0,\cdots,m-x_i-1$), 
as $max\{X\}+t\leq m-1$. 

Thus,
{\small
\begin{eqnarray}\label{equation:nbStates}
                       |Q_i|   \;\leq\;                2^{i-1} (m-x_i)
                       & \leq & 2^{i-1} (m-i), \mbox{ and} \\
      \sum_{i=1}^{r}   |Q_i|   \;\leq\; \sum_{i=1}^{r} 2^{i-1} (m-i)   & =    &(m-r+1) 2^r - m - 1.
\end{eqnarray}}
%The set $Q$ consists of theses states from $\cup^i Q_i$ and $m+1$
%extra states : the final state and $m$ states $<\phi,t>$, where
%$t=0,1,\cdots,m-1$. 
Besides states $Q_i$, $Q$ contains $m$ states $<\emptyset,t>$
($t\in [0..m-1]$) and one final state. 
Thus, $|Q| \leq (m-r+1) 2^r = (w+1) 2^r$.
\end{proof}

%\paragraph{Remark} 
Note that if $\pi$ starts with $\m$, which is always the case for
ordinary spaced seeds, then $X_i \geq i+1$, $i\in[1..r]$, and the
bound of 
(\ref{equation:nbStates}) rewrites to $2^{i-1} (m-i-1)$. 
%instead of $2^{i-1} (m-i)$. 
This results in the same number of states $w 2^r$ as for
the Aho-Corasick automaton \cite{BuhlerKeichSunRECOMB03}.
%
%Note also that 
%the bound of Lemma~\ref{lemma:GnbStates}
%\ref{lemma:GnbStatesConstant}
%is tight, as $S_\pi$ can reach the $\Theta(w \cdot 2^{r})$
%minimal number of states, as the following Lemma states.
The construction of automaton $S_\pi$ is optimal, in the sense that no
two states can be merged in general, as the following Lemma states.

\begin{lemma}\label{lemma:nbStatesWorstCase}
%Let $\pi$ be a spaced seed 
%of span $r+2$ and weight $2$, $\pi$ 
%that
%consists of two "must-match" symbols $\m$ and $r$ joker
%symbols $\j$ between them. 
Consider a spaced seed $\pi$ which consists of two ``must-match''
symbols $\m$ separated by $r$ jokers. 
Then the automaton $S_\pi$ 
%(see lemma \ref{lemma:GnbStates}), 
is reduced, that is
%  \begin{itemize}
%  \item[(i)] 
any non-final state is reachable from the initial state $q_0$, and 
%\item[(ii)] 
any two non-final states $q , q'$ are non-equivalent.
%    i.e. there is a word $w = w(q, q')$ such that exactly one of the states $\psi(q, w), \psi(q', w)$ is a final state.
%  \end{itemize}
\end{lemma}
\begin{proof}
  See appendix \ref{section:AppendixProofnbStatesWorstCase}.
\end{proof}

% -- Constant time spent for each state
%However we have not proved until now that each state is build in
%constant time

A straightforward generation of the transition table of the automaton
$S_\pi$ can be performed in time
$\O(r \cdot w \cdot 2^{r} \cdot |\A| )$. A more complicated algorithm
allows one to reduce the bound  to $\O(w \cdot 2^{r} \cdot
|\A|)$. This algorithm is described in full details in Appendix
\ref{section:AppendixSubsetSeedAutomaton}. Here we summarize it in the
following Lemma.

\begin{lemma}
\label{lemma:GnbStatesConstant}
The transition table of automaton  $S_{\pi}$ can be constructed in
time proportional to its size, which is $\O(w \cdot 2^{r} \cdot |\A|)$.
\end{lemma}
%\begin{proof}
%  See appendix \ref{section:AppendixProofnb????} 
%\end{proof}

% -- #____# seed worst case ( à garder ? à étendre à #__#__# ?)
%The next lemma~\ref{lemma:nbStatesWorstCase} shows that the bound
%$w 2^r$ given in lemma \ref{lemma:GnbStates} cannot be improved
%significantly. Note that lemma~\ref{lemma:nbStatesWorstCase} also
%gives a lower bound for the number of states of the Aho-Corasick
%automaton for spaced seeds, see
%section~\ref{subsection:SeedModelsSeedAutomata}
%and~\cite{BuhlerKeichSunRECOMB03}.  

In the next section, we demonstrate experimentally that on average,
our construction yields a very compact automaton, close to the minimal
one. Together with the general approach of
section~\ref{section:Framework}, this provides a fast algorithm
for computing the sensitivity of subset seeds and, in turn, allows to
perform an efficient design of spaced seeds well-adapted to the
similarity search problem under interest. 

  \section{Experiments}
\label{section:experiments}
Several types of experiments have been performed to test the practical
applicability of the
results of sections~\ref{section:Framework},\ref{section:SubsetSeeds}. 
We focused on DNA similarity search, and set the
alignment alphabet $\A$ to $\{\1,\h,\0\}$ (match, transition, 
transversion). For subset seeds, the seed alphabet $\B$ was set to $
\{\m,\t,\j\}$,  where 
$\m = \{\1\}, \t = \{\1,\h\}, \j = \{\1,\h,\0\}$ (see
Example~\ref{example-subset}). 
%both on spaced and
%subset seeds and their underlying automata 
%using different probability weight transducers and the algorithm of
%section \ref{section:SubsetSeeds}.
%Alignment alphabet $\A$ is $\{\1,\h,\0\}$ (match, transition,
%transversion) and Seed alphabet $\B$ is $ \{\m,\t,\j\}$ where the
%corresponding subsets are the following :
%$\m = \{\1\}, \t = \{\1,\h\}, \j = \{\1,\h,\0\}$.
The weight of a subset seed is computed by assigning weights $1$,
$0.5$ and $0$ to symbols $\m$, $\t$ and $\j$ respectively. 

\subsection{Size of the automaton}
\label{subsection:aut-size}

% -- what is done
We compared the size of the automaton $S_\pi$ defined in
section~\ref{section:SubsetSeeds} and the
Aho-Corasick automaton~\cite{AhoCorasick74}, both for ordinary spaced
seeds (binary seed alphabet) and for subset seeds. The Aho-Corasick
automaton for spaced seeds was constructed as defined in
\cite{BuhlerKeichSunRECOMB03}. For subset seeds, a straightforward
generalization was considered: the Aho-Corasick construction was
applied to the set of alignment fragments matched by the seed. 

Tables~\ref{table:Automaton1}(a) and \ref{table:Automaton1}(b) present
the results for spaced seeds and subset seeds respectively. For each
seed weight $w$, we computed the average number of states ($avg.$ $size$)
of the Aho-Corasick automaton and our automaton $S_\pi$, and reported
the corresponding ratio ($\delta$) with respect to the average number of states of the
minimized automaton. The average was computed over all seeds of span
up to $w+8$ for spaced seeds and all seeds of span up to $w+5$ with two $\t$'s
 for subset seeds.
    \begin{table*}[htb]%
      \vspace{\tbbefore}
      \begin{center}% 
        {\scriptsize%
        \begin{tabular}{c|cc|cc|cc}%
          {\bf Spaced} &
          \multicolumn{2}{|c}{Aho-Corasick}&
          \multicolumn{2}{|c}{$S_\pi$}&
          \multicolumn{1}{|c}{Minimized}\\
          $w$       &
          $avg.$ $size$  &
          $\delta$      &
          $avg.$ $size$  &
          $\delta$      &
          $avg.$ $size$  \\
          \hline
          \hline
          9  & 345.94 & 3.06 & 146.28 & 1.29 & 113.21 \\
          10 & 380.90 & 3.16 & 155.11 & 1.29 & 120.61 \\
          11 & 415.37 & 3.25 & 163.81 & 1.28 & 127.62 \\
          12 & 449.47 & 3.33 & 172.38 & 1.28 & 134.91 \\
          13 & 483,27 & 3.41 & 180.89 & 1.28 & 141.84 \\
        \end{tabular}%
        ~~~
        \begin{tabular}{c|cc|cc|cc}
          {\bf Subset} &
          \multicolumn{2}{|c}{Aho-Corasick}&
          \multicolumn{2}{|c}{$S_\pi$}&
          \multicolumn{1}{|c}{Minimized}\\
          $w$       &
          $avg.$ $size$  &
          $\delta$      &
          $avg.$ $size$  &
          $\delta$      &
          $avg.$ $size$  \\
          \hline
          \hline
          9  & 1900.65 & 15.97  & 167.63 & 1.41 & 119,00\\
          10 & 2103.99 & 16.50  & 177.92 & 1.40 & 127.49\\
          11 & 2306.32 & 16.96  & 188.05 & 1.38 & 135.95\\
          12 & 2507.85 & 17.42  & 198.12 & 1.38 & 144.00\\
          13 & 2709.01 & 17.78  & 208.10 & 1.37 & 152.29\\
        \end{tabular}
        }
        \vspace{\tbinside}\newline
        ~~~~~~~~~(a)~~~~~~~~~~~~~~~~~~~~~~~~~~~~~~~~~~~~~~~~~~~~~~~~~~~~~~~~~~~~~~~~~~~~~~~~~~~~~~~~~~~~~~(b)%\newline
        \caption{\it \label{table:Automaton1} Comparison of the
          average number
          of states of Aho-Corasick automaton, automaton $S_\pi$ of
          section~\ref{section:SubsetSeeds} and minimized automaton}%
      \end{center}%
      \vspace{\tbafter}
    \end{table*}%  
% -- remarks
Interestingly, our automaton turns out to be more compact
than the Aho-Corasick automaton
not only on non-binary alphabets (which was expected), but also on the
binary alphabet (cf Table~\ref{table:Automaton1}(a)). Note that for a
given seed, one can define a 
surjective mapping from the states of the Aho-Corasick
automaton onto the states of our automaton. This implies that our
automaton has {\em always} no more states than the Aho-Corasick
automaton. 
\subsection{Seed Design}
In this part, we considered several probability transducers
to design spaced or subset seeds. The target alignments included all
alignments of length $64$ on alphabet $\{\1,\h,\0\}$. Four
probability transducers have been studied 
%defining the following probabilistic models 
(analogous to those introduced in \cite{BrejovaBrownVinarCPM03}): 
\begin{itemize}
\item $B$: Bernoulli model
\item $DT1$: deterministic probability transducer specifying
  probabilities of $\{\1,\h,\0\}$ at each codon position 
(extension of the $M^{(3)}$ model of~\cite{BrejovaBrownVinarCPM03} to the
three-letter alphabet), 
\item $DT2$: deterministic probability transducer specifying
  probabilities of each of the 27 codon instances $\{\1,\h,\0\}^3$
  (extension of the $M^{(8)}$ model of~\cite{BrejovaBrownVinarCPM03} to the
  three-letter alphabet),
\item $NT$: non-deterministic probability transducer combining four
  copies of $DT2$ specifying four distinct codon conservation levels
  (called HMM model in \cite{BrejovaBrownVinarCPM03}).
\end{itemize}
Models $DT1$, $DT2$ and $NT$ have been trained on alignments resulting
from a pairwise comparison of $40$ bacteria genomes. Details of the
training procedure as well as the resulting parameter values are given
in Appendix~\ref{section:AppendixExperimentalSetup}. 

For each of the four probability transducers, we computed the best
seed of weight $w$ ($w=9,10,11,12$) among two categories: ordinary
spaced seeds of weight $w$ and subset seeds of weight $w$ with two
$\t$. Ordinary spaced seeds were enumerated exhaustively up to a given
span, and for each seed, the sensitivity was computed using the
algorithmic approach of section~\ref{section:Framework} and the seed
automaton construction of section~\ref{section:SubsetSeeds}. Each such
computation took between 10 and 500ms on a Pentium IV 2.4GHz computer
depending on the seed weight/span and the model used. In each
experiment, the most sensitive seed found has been kept. The results
are presented in Tables~\ref{table:Bernoulli}-\ref{table:HMM}.

\begin{table*}[htb]%
  \vspace{\tbbefore}
  \begin{center}% 
    \vspace{\tbinside}
            {%\small
    \begin{tabular}{c|lc|lc}
      {$w$} &
      {~~spaced seeds} &
      {~~Sens.} & 
      {~~subset seeds, two $\t$} &
      {~~Sens.} \\
      \hline
      \hline
      9  &~\mbox{\tt \#\#\#\_\_\_\#\_\#\_\#\#\_\#\#}~&       0.4183
      &~\mbox{\tt \#\#\#\_\#\_\_\#@\#\_@\#\#}~&           0.4443\\
      10 &~\mbox{\tt \#\#\_\#\#\_\_\_\#\#\_\#\_\#\#\#}~&     0.2876
      &~\mbox{\tt \#\#\#\_@\#\_@\#\_\#\_\#\#\#}~&         0.3077\\
      11 &~\mbox{\tt \#\#\#\_\#\#\#\_\#\_\_\#\_\#\#\#}~&     0.1906
      &~\mbox{\tt \#\#@\#\_\_\#\#\_\#\_\#\_@\#\#\#}~&     0.2056\\
      12 &~\mbox{\tt \#\#\#\_\#\_\#\#\_\#\_\_\#\#\_\#\#\#}~& 0.1375
      &~\mbox{\tt \#\#@\#\_\#\_\#\#\_\_\#@\_\#\#\#\#}~&   0.1481\\
    \end{tabular}
   }%
    \caption{\it \label{table:Bernoulli} Best seeds and their sensitivity for probability transducer $B$}
  \end{center}%
  \vspace{\tbafter}
\end{table*}%  
\begin{table*}[htb]%
  \vspace{\tbbefore}
    \begin{center}
      \vspace{\tbinside}
             {%\small
	       \begin{tabular}{c|lc|lc}
		 {$w$} &
		 {~~spaced seeds} &
		 {~~Sens.} & 
		 {~~subset seeds, two $\t$} &
		 {~~Sens.} \\
		 \hline 
		 \hline
		 9 & \mbox{\tt \#\#\#\_\_\_\#\#\_\#\#\_\#\# }~& 0.4350
		 & \mbox{\tt \#\#@\_\_\_\#\#\_\#\#\_\#\#@ }~& 0.4456\\
		 10 & \mbox{\tt \#\#\_\#\#\_\_\_\_\#\#\_\#\#\_\#\# }~& 0.3106
		 & \mbox{\tt \#\#\_\#\#\_\_\_@\#\#\_\#\#@\# }~& 0.3173\\
		 11 & \mbox{\tt \#\#\_\#\#\_\_\_\_\#\#\_\#\#\_\#\#\# }~& 0.2126
		 & \mbox{\tt \#\#@\#@\_\#\#\_\#\#\_\_\#\#\# }~& 0.2173\\
		 12 & \mbox{\tt \#\#\_\#\#\_\_\_\_\#\#\_\#\#\_\#\#\#\# }~& 0.1418
		 & \mbox{\tt \#\#\_@\#\#\#\_\_\#\#\_\#\#@\#\# }~& 0.1477\\
	     \end{tabular}}%
	     \caption{\it \label{table:M3} Best seeds and their sensitivity for probability transducer $DT1$}
    \end{center}%
\end{table*}%  
%
%The first trained model is a $M3$ ~\cite{BrejovaBrownVinarCPM03}
%one. It consists of a three state weighed automaton. This weighed
%transducer is fitted to simulate more accurately regularities involved
%in DNA coding regions.
%
\vspace{-5mm}
\begin{table*}[htb]%
  \vspace{\tbbefore}
    \begin{center}
      \vspace{\tbbefore}
      \vspace{\tbinside}
	     {%small
               \begin{tabular}{c|lc|lc}
                 {$w$} &
                 {~~spaced seeds} &
                 {~~Sens.} & 
                 {~~subset seeds, two $\t$} &
                 {~~Sens.}\\
                 \hline 
                 \hline
		 9  &~\mbox{\tt \#\_\#\#\_\_\_\_\#\#\_\#\#\_\#\#}~&         0.5121
		 &~\mbox{\tt \#\_\#@\_\#\#\_@\_\_\#\#\_\#\#}~&           0.5323\\
		 10 &~\mbox{\tt \#\#\_\#\#\_\#\#\_\_\_\_\#\#\_\#\#}~&       0.3847
		 &~\mbox{\tt \#\#\_@\#\_\#\#\_\_@\_\#\#\_\#\#}~&         0.4011\\
		 11 &~\mbox{\tt \#\#\_\#\#\_\_\#\_\#\_\_\_\#\_\#\#\_\#\#}~& 0.2813
		 &~\mbox{\tt \#\#\_\#\#\_@\#\_\#\_\_\_\#\_\#@\_\#\#}~&   0.2931\\
		 12 &~\mbox{\tt \#\#\_\#\#\_\#\#\_\#\_\_\_\#\_\#\#\_\#\#}~& 0.1972
		 &~\mbox{\tt \#\#\_\#\#\_\#@\_\#\#\_@\_\_\#\#\_\#\#}~&   0.2047\\
               \end{tabular}
             }%
	     \caption{\it \label{table:M14} Best seeds and their sensitivity for probability transducer $DT2$} 
    \end{center}%
    \vspace{\tbafter}
\end{table*}%  
%
%Second trained model is the $M14$ one ($M8$
%~\cite{BrejovaBrownVinarCPM03} equivalent). This weighed transducer
%can be seen as a more accurate $M3$ model that takes into account
%some dependencies between generated letters on each codon.
%
\begin{table*}[htb]%
  \vspace{\tbbefore}
  \begin{center}
    \vspace{\tbinside}
           %{\scriptsize%\small
             \begin{tabular}{c|lc|lc}
               {$w$} &
               {~~spaced seeds} &
               {~~Sens.} & 
               {~~subset seeds, two $\t$} &
               {~~Sens.} \\              
               \hline 
               \hline
	       9  &~\mbox{\tt \#\#\_\#\#\_\#\#\_\_\_\_\#\#\_\#}  & 0.5253
	       &~\mbox{\tt \#\#\_@@\_\#\#\_\_\_\_\#\#\_\#\#} & 0.5420\\   
	       10 &~\mbox{\tt \#\#\_\#\#\_\_\_\_\#\#\_\#\#\_\#\#} & 0.4123        
	       &~\mbox{\tt \#\#\_\#\#\_\_\_\_\#\#\_@@\_\#\#\_\#} & 0.4190\\        
	       11 &~\mbox{\tt \#\#\_\#\#\_\_\_\_\#\#\_\#\#\_\#\#\_\#} & 0.3112        
	       &~\mbox{\tt \#\#\_\#\#\_\_\_\_\#\#\_@@\_\#\#\_\#\#} & 0.3219\\       
	       12 &~\mbox{\tt \#\#\_\#\#\_\_\_\_\#\#\_\#\#\_\#\#\_\#\#} & 0.2349        
	       &~\mbox{\tt \#\#\_\#\#\_\_\_\_\#\#\_@@\_\#\#\_\#\#\_\#} & 0.2412\\
             \end{tabular}
           %}%
	   \caption{\it \label{table:HMM} Best seeds and their sensitivity for probability transducer $NT$} 
  \end{center}%
  \vspace{\tbafter}
\end{table*}%  
%
%Third trained model is an $HMM$ equivalent. It consists in 4 $M14$
%models with non deterministic transitions between each of the 4 $M14$
%first states. Those 4 classes have been selected according to the DNA
%level of similarity as in ~\cite{BrejovaBrownVinarCPM03}.
%%
In all cases, subset seeds yield a better sensitivity than ordinary
spaced seeds. The sensitivity increment varies up to 0.04 which is a
notable increase. As shown in \cite{NoeKucherovBMC04}, the gain in using subset
seeds increases substantially when the transition probability is
greater than the inversion probability, which is very often the case
in related genomes.
\subsection{Comparative performance of spaced and subset seeds}
We performed a series of whole genome comparisons in order to compare
the performance of designed spaced and subset seeds. Eight complete bacterial
genomes%
\footnote{
NC\_000907.fna,
NC\_002662.fna,
NC\_003317.fna,
NC\_003454.fna,
NC\_004113.fna,
NC\_001263.fna,
NC\_003112.fna obtained from NCBI}
have been processed against 
each other using the YASS software \cite{NoeKucherovBMC04}. Each
comparison was done twice: one with a spaced seed and another with a
subset seed of the same weight. 

The threshold E-value for the output alignments was set to $10$, and
for each comparison, 
the number of alignments with E-value smaller than $10^{-3}$ found by
each seed, and the number of exclusive alignments were reported.
By ``exclusive alignment'' we mean any alignment of E-value less than
$10^{-3}$ that does not share a common part (do not overlap in both
compared sequences) with any alignment found by another seed.
To take into account a possible bias caused by splitting alignments
into smaller ones (X-drop effect), we also computed the total length
of exclusive alignments. 
Table~\ref{table:seedexperiments} summarizes these
experiments for weights 9 and 10 and the $DT2$ and $NT$ probabilistic
models. Each line corresponds to a seed given in
Table~\ref{table:M14} or Table~\ref{table:HMM}, depending on the
indicated probabilistic model.
\begin{table*}[htb]%
  \vspace{\tbbefore}
  \begin{center}%
    \vspace{\tbinside}
    {\scriptsize
      \begin{tabular}{c|cccc} 
        $seed$  &
        $time$  &
        $\# align$     & 
        $\# ex. align$ &
        $ex.$~$align$~$length$\\
        \hline
        \hline
        $DT2$, $w=9$, spaced seed        &  15:14 & 19101& 1583 & 130512 \\ 
        $DT2$, $w=9$, subset seed, two $\t$ &  14:01 & 20127 & 1686 & 141560 \\
        \hline
        $DT2$, $w=10$, spaced seed        &  8:45 & 18284 & 1105 & 10174 \\
        $DT2$, $w=10$, subset seed, two $\t$  &  8:27 & 18521 & 1351 & 12213 \\
        \hline
        $NT$, $w=9$, spaced seed        &  42:23 & 20490 & 1212 & 136049 \\
        $NT$, $w=9$, subset seed, two $\t$  &  41:58 & 21305 & 1497 & 150127 \\
        \hline
        $NT$, $w=10$, spaced seed       &  11:45 & 19750 & 942 & 85208 \\
        $NT$, $w=10$, subset seed, two $\t$  &  10:31 & 21652 & 1167 & 91240 \\
        \hline
      \end{tabular}%
      }%
    \caption{\it \label{table:seedexperiments} Comparative test of
    subset seeds vs spaced seeds. Reported execution times (min:sec) were obtained on a Pentium IV 2.4GHz computer.}
    \end{center}%
  \vspace{\tbafter}
\end{table*}%  
%
%The table implies that best subset seeds detect 2 to 5 percent more of
%significant alignments missed by best spaced seeds of the same weight. 
In all cases, best subset seeds detect from 1\% to 8\% more significant alignments compared to best spaced seeds of same weight.
  \section{Discussion}
\label{section:discussion}

We introduced a general framework for computing the seed sensitivity
for various similarity search settings.  The approach 
can be seen as a generalization of methods of
\cite{BuhlerKeichSunRECOMB03,BrejovaBrownVinarJBCB04}  
in that it allows to obtain algorithms with the same worst-case
complexity bounds as those proposed in these papers, but also allows
to obtain efficient algorithms for new formulations of the seed
sensitivity problem. This versatility is achieved by distinguishing
and treating separately the three ingredients of the seed sensitivity
problem: a set of target alignments, an associated probability
distributions, and a seed model.

We then studied a new concept of {\em subset seeds} which represents an interesting
compromise between the efficiency of spaced seeds and the flexibility
of vector seeds. For this type of seeds, we defined an automaton with
$\O(w2^r)$ states regardless of the size of the alignment alphabet,
and showed that its transition table can be constructed in time
$\O(w2^r|\A|)$. Projected to the case of spaced seeds, this
construction gives the same worst-case bound as the Aho-Corasick
automaton of \cite{BuhlerKeichSunRECOMB03}, but results in a smaller
number of states in practice. Different experiments we have done
confirm the practical efficiency of the whole method, both at the level
of computing sensitivity for designing good seeds, as well as using
those seeds for DNA similarity search. 

As far as the future work is concerned, it would be interesting to
study the design of efficient spaced seeds for protein sequence
search (see \cite{BrownTCBB05}), as well as to combine spaced seeds
with other techniques such as 
% daughter seeds \cite{CsurosMa04}.
seed families
\cite{PatternHunter04,BuhlerRECOMB04,KucherovNoeRoytbergTCBB05} or the
group hit criterion \cite{NoeKucherovBMC04}. 

\paragraph{Acknowledgements}
%\noindent 
%{\bf Acknowledgements}
%The work of M.R. has been partially supported by grants RFFI
%03-04-4949369 of the Russian Foundation for Basic Research, and
%20/2002 from the Russian Ministry of Science and Technology 
%and from NWO (Netherland Science Foundation). 
%
G.~Kucherov and L.~No\'e have been supported
by the {\em ACI IMPBio} of the French Ministry of Research. 
A part of this work has been done during a stay of M.~Roytberg at
LORIA, Nancy, supported by INRIA. 
M.Roytberg has been also supported by the Russian Foundation for
Basic Research (projects 03-04-49469, 02-07-90412) and by grants
from the RF Ministry of Industry, Science and Technology (20/2002,
5/2003) and NWO (Netherlands Science Foundation).

  {\scriptsize
    \bibliographystyle{acm}
    \bibliography{paper.bib}
  }
  \newpage
  \onecolumn
  \appendix
\section{Proof of Lemma \ref{lemma:nbStatesWorstCase}}
\label{section:AppendixProofnbStatesWorstCase}

%\begin{lemma}\label{lemma:Appendix02}
  Let $\pi = \#-^r\#$ be a spaced seed of span $r+2$ and weight
  $2$. We prove that the automaton $S_\pi$ (see Lemma~\ref{lemma:Gpi}) is reduced, i.e.
  \begin{itemize}
  \item[(i)] all its non-final states are reachable from the initial state $<\emptyset , 0>$;
  \item[(ii)] any two non-final states $q , q'$ are non-equivalent, 
    i.e. there is a word $w = w(q, q')$ such that exactly one of the states $\psi(q, w), \psi(q', w)$ is a final state. 
  \end{itemize}
%\end{lemma}

%\begin{proof}~\\

%We prove that 
%  \begin{itemize}
%  \item[A]. 
(i) Let $q = <X, t>$ be a state of the automaton 
$S_\pi$, and let $X =\{x_1,\ldots, x_k\}$ and  $ x_1 < \cdots <  x_k$. 
Obviously, $x_k + t < r+2$. Let $s\in \{0, 1\}^*$ be an
alignment word of length $x_k$ such that for  all 
$i \in [1, x_k], \;   s_i = \1 \mbox{~iff~} \exists j \in [1,k], \quad i = x_k - x_j + 1$. 
Note, that, $\pi_1=\#$, therefore $1 \notin X $ and $s_{x_k}=0$. 
Finally, $\psi (<\phi, 0>, s \cdot 1^t) = q$.

%  \item[B]. 
(ii)
Let $q_1 = <X_1, t_1>$ and $q_2 = <X_2, t_2>$ be non-final
    states of $S_\pi$. Let $X_1 = \{y_1, \ldots, y_a\}, X_2 = \{z_1,\ldots, z_b\}$, and
    $y_1 < \cdots < y_a$,  $z_1 < \cdots < z_b$.

Assume that $max \{ X_1 \} + t_1 > max \{ X_2 \} + t_2$
    and let $d = (r+2) - (max \{ X_1 \} + t_1)$. Obviously,
    $\psi(q_1 ,1^d)$ is a final state, and $\psi(q_2 ,1^d)$ is not.
    Now assume that $max \{ X_1 \} + t_1 = max \{ X_2 \} + t_2$.
For a set $X \subseteq \{1,\ldots , r+1\}$ and a number $t$, define
a set $X\{t\}$ by $X\{t\} = \{v+t | v \in X  \mbox{~and~} v+t < r+2 \}$.
Let $g = max \{ v | ( v+t_1 \in X_1 \mbox{~ and~} v+t_2 \notin X_2) \mbox{~or~}    ( v+t_2 \in X_2  \mbox{~and~} v+t_1 \notin X_1)  \}$
and let  $d = r+1 - g$ . Then $\psi( q_1 ,0^{d} \cdot 1)$ is a final state and $\psi( q_2 , 0^{d} \cdot 1)$ is not or vice versa.
This completes the proof.

%  \end{itemize}
  %The proof is accomplished.
%\end{proof}

% -- Proof Subset seed automaton w2^r build procedure

\section{Subset seed automaton}
\label{section:AppendixSubsetSeedAutomaton}
Let $\pi$ be a subset seed of $\#$-weight $w$ and span $s$, and $r =
s-w$ be the number of non-$\#$ positions. We define a DFA $S_\pi$
recognizing all words of $\A^*$ matched by $\pi$ (see definition of
section~\ref{subsection:SubsetSeeds}). The transition table of $S_\pi$
is stored in an array such that each element describes a state $<X,t>$ of $S_\pi$.
Now we define
\begin{itemize}
\item[1.] how to compute the array index $Ind(q)$ of a state
  $q=<X,t>$, 
\item[2.] how to compute values $\psi(q,a)$ given a state $q$ and
    a letter $a \in \A$.
\end{itemize}

% -- Indexes of states if stored inside a table
\subsection{Encoding state indexes}
\label{section:AppendixStateIndices}

We will need some notation. 
Let $L= \{ l_1,\ldots,l_r \}$ be a set of all non-$\#$ positions in
$\pi$ ($l_1 < l_2 < \cdots < l_r$). For a subset $X \subseteq L$, let
$v(X)=v_1 \ldots v_r \in \{0,1\}^r$ be a binary vector such that $v_i
= 1$ 
%\iff 
iff $l_i \in X$. Let further $n(X)$ be the integer corresponding to
the binary representation $v(X)$ (read from left to right):

\[
n(X) = \sum_{j=1}^{r} 2^{j-1} \cdot v_j.
\]

% States $\{<X,t>\}$ are ordered, first by $t$, then by $n(X)$.

Define $p(t) = max \{ p\: |\: l_p < m - t \}$. Informally, for a
given non-final state $<X,t>$, $X$ can only be a subset of
$\{l_1,\ldots, l_{p(t)}\}$. This implies that $n(X) < 2^{p(t)}$.
Then, the index of a given state  $\{<X,t>\}$ in the array is defined by
\[
  Ind(<X,t>)=n(X)+2^{p(t)}.
\]
This implies that the worst-case size of the array is no more than
$w2^r$ (the proof is similar to the proof of
Lemma~\ref{lemma:GnbStates}). 
%
%In practice, an inductive generation of reachable states greatly improves the
%efficiency.

\subsection{Computing transition function $\psi(q,a)$}
\label{section:AppendixTransitionFunction}

We compute values $\psi(<X,t>,a)$ based on already computed values $\psi(<X',t>,a)$.
Let $q=<X,t>$ be a non-final and reachable state of $S_\pi$,  where
$X=\{l_{1},\ldots,l_{k}\}$ with $l_{1} < l_{2} \cdots < l_{k}$ and 
$k \leq r$.  
Let $X' = X \setminus \{l_k\} = \{l_{1},\ldots,l_{k-1}\}$ and
$q'=<X',t>$. Then the following lemma holds.

% -- reachability proof
\begin{lemma}\label{lemma:TransitionFunctionPrecedence}
If $q=<X,t>$ is reachable, then $q'=<X',t>$ is reachable and has been processed before.
\end{lemma}
  
\begin{proof}
First prove that $<X',t>$ is reachable. 
 If $<X,t>$ is reachable, then  $<X,0>$  is reachable due to the
 definition of transition function for $t > 0$.  Thus, one can find at least one
 sequence $S \in \A^{l_k}$ such that $\forall i \in [1..r]$,   
$l_i \in X$   
%\iff  
iff
$\pi_{1} \cdots \pi_{l_i}$ matches $S_{l_k-l_i+1}
 \cdots  S_{l_k}$.
For such a sequence $S$, one can find a word $S' =
S_{l_{k}-l_{k-1}+1} \cdots S_{l_{k}}$ which reaches state $<X',0>$. 
To conclude, if there exists a word $S\cdot 1^t$ that reaches 
the state $<X,t>$, there also exists a word 
$S'\cdot 1^t$ that reaches $<X',t>$. 

Note that as $|S' \cdot 1^t| <  |S \cdot 1^t| $, then a breadth-first
computation of states of $S_\pi$ always processes state $<X',t>$
before $<X,t>$.
\end{proof}%
  Now we present how to compute values $\psi(<X,t>,a)$
  from values $\psi(<X',t>,a)$. This is done by
  Algorithm~\ref{algorithm:subsetseedautomaton} shown below, that we comment on now. 
  Due to implementation choices, we represent a state $q$ as
  triple $q =  \langle X,k_X,t \rangle$, where $k_X = max\{ i | l_i \in  X\}$.
  Note first that if $a  = \1$, the transition function $\psi(q,a)$
  can be computed in constant time due to its definition (part a. of Algorithm~\ref{algorithm:subsetseedautomaton}).
If $a  \neq \1$, we have to
\vspace{-2mm}%
  \begin{itemize}
  \item[1.] retrieve the index of $q'$ given $q =  \langle X,k_X,t \rangle$  (part c. of  Algorithm~\ref{algorithm:subsetseedautomaton}), %
  \vspace{-2mm}%
  \item[2.] compute $\psi(\langle X,k_X,t \rangle,a \neq \1)$ given
    $\psi(\langle X',k_{X'},t \rangle,a \neq \1)$
    value. (part d. of Algorithm~\ref{algorithm:subsetseedautomaton}) 
  \end{itemize}
  \vspace{-2mm}%
  \paragraph{\rm 1.} Note first that $Ind(\langle X,k_X,t \rangle) =
  Ind(\langle X',k_{X'},t \rangle) - 2^{k_X}$, which can be computed
  in constant time since $k_X$ is explicitly stored in the current state.
  \vspace{-2mm}%
  \paragraph{\rm 2.}  Let 
  \begin{eqnarray*}
    V_X (k,t,a \neq \1) & = & \begin{cases}  
      l_i & \mbox{~if~} l_i = l_k + t + 1  \mathrm{~and~} \; a  \mathrm{~matches~} \pi_{l_i} \\  
      \emptyset & \mbox{~otherwise~} \\
    \end{cases}\\ 
  \end{eqnarray*}  
  and 
  \vspace{-2mm}
  \begin{eqnarray*}
    V_k (k,t,a \neq \1) & = & \begin{cases}  
      i & \mbox{~if~} l_i = l_k + t + 1  \mathrm{~and~} \; a  \mathrm{~matches~} \pi_{l_i} \\  
      0 & \mbox{~otherwise~} \\
    \end{cases}\\ 
  \end{eqnarray*} 

  Tables $V_X (k,t,a)$ and $V_k (k,t,a)$ can be
  precomputed in time and space $\mathcal{O}(|\A| \cdot m^2)$.
  Let $\psi(\langle X,k_X,t \rangle,a) = \langle Y,k_Y,0
  \rangle$ and  $\psi(\langle X',k_{X'},t \rangle,a) =
  \langle Y',k_{Y'},0 \rangle$. The set $Y$ differs from $Y'$ at most
  with one element. This element can be computed in constant time using
  tables $V_X,V_k$. Namely $Y = Y' \cup V_X(k_X,t,a)$ and $k_Y =
  max(k_{Y'},V_k(k_X,t,a))$.

  Note that a final situation arises when $X = \emptyset$. (part b. of Algorithm~\ref{algorithm:subsetseedautomaton}).
  One also has to compute two tables $U_X,U_k$ defined as:
  \begin{eqnarray*}
    U_X (t,a \neq \1) & = & \cup \{  x    | x \leq t + 1 \mathrm{~and~} a \mathrm{~matches~} \pi_{x}\}\\
    U_k (t,a \neq \1) & = & max \{ x    | x \leq t + 1 \mathrm{~and~} a \mathrm{~matches~} \pi_{x}\}\\
  \end{eqnarray*} 
  
  \begin{lemma}\label{lemma:TransitionFunctionConstantTime}
    The transition function $\psi(q,a)$ can be computed in constant time
    for every reachable state $q$ and every $a \in \A$.
  \end{lemma}

  \def\State#1{ \langle {#1} \rangle } 

  \begin{algorithm2e}
    \label{algorithm:subsetseedautomaton}
    \caption{ $S_{\pi}$ computation }
    \AlgData{a seed $\pi$ of span $m$, $'\#'$-weight $w$, and number of jokers $r = m - w$}
    \AlgResult{an automaton $S_\pi = <Q,q_0,q_F,\mathcal{A},\psi>$}      
        
    $Q.add(q_F)$\;
    $q_0 \leftarrow \State{X=\emptyset,k=0,t=0} $ \;
    $Q.add(q_0)$\;
    $queue.push(q_0)$\; 
    \While{$queue \neq \emptyset$}{
      $\State{X,k_X,t_X} = queue.pop()$\;
      \For{$a \in \mathcal{A}$}{
        \emph{/* compute $\psi(<X,t_X>,a) = \State{Y,k_Y,t_Y}$ */ }\\
        \eIf{$a = \1$}{
            $t_{Y} \leftarrow t_{X} + 1 $\;
\nlset{a} $k_{Y} \leftarrow k_{X}$\;
            $Y \leftarrow X$\;
        }{
          \eIf{$X = \emptyset$}{             
\nlset{b} $Y \leftarrow  U_{X}(t_X,a)$\;
            $k_{Y} \leftarrow U_{k}(t_X,a)$\; 
          }{
            \emph{/* use already processed $\psi(<X',t_{X'}>,a)$ \ldots */ }\\
\nlset{c}   $X' \leftarrow X \; \backslash\;  \{ l_{k_{X}} \}$\;
            $\State{Y',k_{Y'},t_{Y'}} \leftarrow \psi(<X',t>,a)$\; 
            \emph{/* \ldots to compute $\psi(<X,t_X>,a)$  */ }\\
            
\nlset{d} $k_{Y} \leftarrow max\big(k_{Y'},V_{k}(k_X,t_X,a)\big)$\;
            $Y \leftarrow  Y' \cup V_{X}(k_X,t_X,a)$\;
          }
          $t_{Y} \leftarrow 0$\;
        }
        \eIf{$L[k_Y] + t_Y \geq m$}{
          \emph{/* $<Y,t_Y>$ is a final state */}\\
          $\psi(<X,t_X>,a) \leftarrow q_{F}$\;
        }{
	\If{$\State{Y,k_Y,t_Y} \notin Q$}{
              $Q.add(\State{Y,k_Y,t_Y})$\;
              $queue.push(\State{Y,k_Y,t_Y})$\;
           }
          $\psi(<X,t_X>,a) \leftarrow \State{Y,k_Y,t_Y}$\;
        }
      }
    }  
  \end{algorithm2e}  \newpage

%\section{Details of Experimental Setup}
\section{Training probability transducers}
\label{section:AppendixExperimentalSetup}
%\subsection{Training probability transducers}
%\label{section:AppendixExperimentalSetup:A}
We selected 40 bacterial complete genomes from NCBI:
{\footnotesize
NC\_000117.fna,
NC\_000907.fna,
NC\_000909.fna,
NC\_000922.fna,
NC\_000962.fna,
NC\_001263.fna,
NC\_001318.fna,
NC\_002162.fna,
NC\_002488.fna,
NC\_002505.fna,
NC\_002516.fna,
NC\_002662.fna,
NC\_002678.fna,
NC\_002696.fna,
NC\_002737.fna,
NC\_002927.fna,
NC\_003037.fna,
NC\_003062.fna,
NC\_003112.fna,
NC\_003210.fna,
NC\_003295.fna,
NC\_003317.fna,
NC\_003454.fna,
NC\_003551.fna,
NC\_003869.fna,
NC\_003995.fna,
NC\_004113.fna,
NC\_004307.fna,
NC\_004342.fna,
NC\_004551.fna,
NC\_004631.fna,
NC\_004668.fna,
NC\_004757.fna,
NC\_005027.fna,
NC\_005061.fna,
NC\_005085.fna,
NC\_005125.fna,
NC\_005213.fna,
NC\_005303.fna,
NC\_005363.fna 
}.

YASS \cite{NoeKucherovBMC04} has been run on each pair of genomes to detect alignments with
E-value at most $10^{-3}$. Resulting ungapped regions of length $64$
or more have been used to train models $DT1$, $DT2$ and $NT$ by the maximal 
likelihood criterion.
Table~\ref{table:M3Model} gives the $\rho$ function of the
probability transducer $DT1$, that specifies the probabilities of
match ($\1$), transition ($\h$) and transversion ($\0$) at each codon
position. 
\vspace{-10mm}
\begin{table*}[htb]\center
          {\scriptsize%\small
  \begin{tabular}{c|lll}
    $ a : $ & ~~~$\0$   & ~~~$\h$   & ~~~$\1$  \\
    \hline
    \hline
    $\rho(q_0,a,q_1)$ ~& 0.2398 & 0.2945 & 0.4657 \\
    $\rho(q_1,a,q_2)$ ~& 0.1351 & 0.1526 & 0.7123 \\
    $\rho(q_2,a,q_0)$ ~& 0.1362 & 0.1489 & 0.7150 \\
  \end{tabular}
  }%
  ~~\includegraphics[width=2cm]{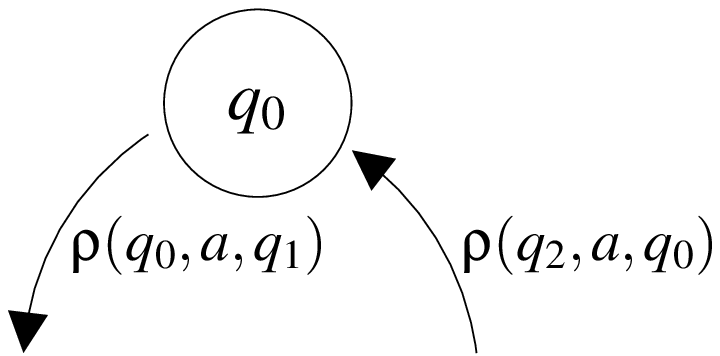} 
  \caption{\it \label{table:M3Model} Parameters of the $DT1$ model}
%  \vspace{-10mm}
\end{table*}

Table \ref{table:M14Model} specifies the probability of each codon
instance $a_1 a_2 a_3 \in \A^3$, used to define the probability
transducer $DT2$. 
%$M14$ parameters (Table \ref{table:M14Model}) is
%implicitly given by the probability of each word $a_1 a_2 a_3 \in \A^3$. 
\begin{table*}[htb]\center
  {\scriptsize%
  \begin{tabular}{c|lllllllll}
    $a_1 \backslash a_2a_3 : $& $\q{00}$ & $\q{0h}$ & $\q{01}$ & $\q{h0}$ &  $\q{hh}$ & $\q{h1}$ & $\q{10}$  & $\q{1h}$   & $\q{11}$ \\
    \hline
    \hline
    $\0$ & 0.01089 &   0.01329 &   0.01311 &   0.01107 &   0.00924 &  0.01144 &   0.01887 &  0.01946 &  0.03106\\
    $\h$ & 0.01022 &   0.00984 &   0.01093 &   0.00956 &   0.01025 &  0.01294 &   0.02155 &  0.02552 &  0.03983\\
    $\1$ & 0.02083 &   0.02158 &   0.02554 &   0.02537 &   0.02604 &  0.03776 &   0.11298 &  0.16165 &  0.27915\\
  \end{tabular}}%
  \caption{\it \label{table:M14Model} Probability of each codon
    instance specified by the $DT2$ model}
\end{table*}

Finally, Table \ref{table:HMMModel} specifies the probability
transducer $NT$ by specifying the four $DT2$ models together with
transition probabilities between the initial states of each of these
models. 
%model parametrisation () is given by four $M14$ models and a
%transition table for each first state of each $M14$ model.
\begin{table*}[htb]\center
  {\scriptsize
  \begin{tabular}{c|llll}
    $Pr(q_i \rightarrow q_j)$ & $j=0$ & $~~1$ & $~~2$ & $~~3$ \\
    \hline
    \hline
    $i=0$     & 0.9053 & 0.0947 & 0      & 0\\
    $~~~~~~1$ & 0.1799 & 0.6963 & 0.1238 & 0\\
    $~~~~~~2$ & 0      & 0.2131 & 0.6959 & 0.0910\\
    $~~~~~~3$  & 0.0699 & 0.0413 & 0.1287 & 0.7601\\
  \end{tabular}

  \begin{tabular}{c|lllllllll}
    $a_1  \backslash a_2 a_3 : $ &$\q{00}$ & $\q{0h}$ & $\q{01}$ & $\q{h0}$ &  $\q{hh}$ & $\q{h1}$ & $\q{10}$  & $\q{1h}$   & $\q{11}$ \\
    \hline
    \hline
    %$~~~~~~~\0$ &	0.01892 & 0.01345 & 0.02855 & 0.01203 & 0.01018 & 0.02104 & 0.02940 & 0.03017 & 0.09154 \\
    %$q_0 :\h$ & 	0.01281 & 0.00973 & 0.01950 & 0.00921 & 0.00639 & 0.01710 & 0.02108 & 0.01853 & 0.09715 \\
    %$~~~~~~~\1$ &	0.02120 & 0.01801 & 0.04802 & 0.01716 & 0.01422 & 0.05317 & 0.07983 & 0.08940 & 0.19221 \\
    %\hline
    %$~~~~~~~\0$ &	0.01395 & 0.01007 & 0.02127 & 0.00198 & 0.00595 & 0.01790 & 0.01995 & 0.01519 & 0.09714 \\ 
    %$q_1 :\h$ &         0.00978 & 0.00726 & 0.01781 & 0.00644 & 0.00415 & 0.01462 & 0.01437 & 0.01269 & 0.10545 \\
    %$~~~~~~~\1$ &	0.02149 & 0.01586 & 0.05348 & 0.01734 & 0.01186 & 0.05405 & 0.08792 & 0.09274 & 0.24919 \\
    %\hline
    %$~~~~~~~\0$ &    	0.00985 & 0.00572 & 0.01638 & 0.00683 & 0.00399 & 0.01271 & 0.01510 & 0.01179 & 0.09681 \\ 
    %$q_2 :\h$  &        0.00648 & 0.00504 & 0.01613 & 0.00572 & 0.00251 & 0.01170 & 0.01182 & 0.00896 & 0.11057 \\
    %$~~~~~~~\1$ &	0.01764 & 0.01513 & 0.05149 & 0.01319 & 0.00986 & 0.05504 & 0.08301 & 0.08628 & 0.31025 \\
    %\hline 
    %$~~~~~~~\0$ &       0.00694 & 0.00492 & 0.01023 & 0.00426 & 0.00264 & 0.01066 & 0.01183 & 0.00938 & 0.08264 \\
    %$q_3 :\h$ 	&       0.00473 & 0.00432 & 0.00901 & 0.00529 & 0.00123 & 0.01107 & 0.00899 & 0.00720 & 0.12063 \\
    %$~~~~~~~\1$	&	0.01248 & 0.01027 & 0.04383 & 0.01028 & 0.00892 & 0.05375 & 0.07112 & 0.07409 & 0.39929 \\
    $~~~~~~~\0$ & 0.01577 & 0.01742 & 0.01440 & 0.01511 & 0.01215 & 0.01135 & 0.02502 & 0.02353 & 0.02786 \\
    $q_0 :\h$   & 0.01478 & 0.01365 & 0.01266 & 0.01348 & 0.01324 & 0.01346 & 0.02815 & 0.02981 & 0.03442 \\
    $~~~~~~~\1$ & 0.02701 & 0.02838 & 0.02600 & 0.03429 & 0.03158 & 0.03406 & 0.12973 & 0.17461 & 0.17809 \\
    \hline
    $~~~~~~~\0$ & 0.00962 & 0.01241 & 0.01501 & 0.00891 & 0.00753 & 0.01247 & 0.01791 & 0.01841 & 0.03530 \\
    $q_1 :\h$   & 0.00818 & 0.00766 & 0.01115 & 0.00738 & 0.00952 & 0.01353 & 0.01828 & 0.02978 & 0.04405 \\
    $~~~~~~~\1$ & 0.01946 & 0.01682 & 0.02344 & 0.02456 & 0.02668 & 0.03890 & 0.12113 & 0.18170 & 0.26020 \\
    \hline
    $~~~~~~~\0$ & 0.00406 & 0.00692 & 0.00954 & 0.00501 & 0.00372 & 0.00841 & 0.01034 & 0.01129 & 0.03430 \\
    $q_2 :\h$   & 0.00391 & 0.00396 & 0.00758 & 0.00364 & 0.00707 & 0.01473 & 0.01288 & 0.01975 & 0.05058 \\
    $~~~~~~~\1$ & 0.01250 & 0.01627 & 0.02416 & 0.01419 & 0.02071 & 0.04427 & 0.10014 & 0.15311 & 0.39698 \\ 
    \hline
    $~~~~~~~\0$ & 0.00302 & 0.00267 & 0.00560 & 0.00289 & 0.00249 & 0.00807 & 0.00740 & 0.00710 & 0.03195 \\
    $q_3 :\h$   & 0.00297 & 0.00261 & 0.00355 & 0.00299 & 0.00271 & 0.00935 & 0.00924 & 0.01148 & 0.04296 \\
    $~~~~~~~\1$ & 0.01035 & 0.01125 & 0.02204 & 0.00930 & 0.01289 & 0.04235 & 0.05304 & 0.08163 & 0.59810 \\
  \end{tabular}%
  }%
  \caption{\it \label{table:HMMModel} Probabilities specified by the
    $NT$ model}
\end{table*}

%\subsection{}
%\label{section:AppendixExperimentalSetup:B}
%We selected 8 bacterial complete genome from NCBI:
%{\footnotesize
%NC\_000907.fna,

%NC\_002662.fna,
%NC\_003317.fna,
%NC\_003454.fna,
%NC\_004113.fna,
%NC\_001263.fna,
%NC\_003112.fna
%}.
%YASS has been run on each pair of sequences to detect all alignments with
%E-value $10$, but only alignments with E-value $< 10^{-3}$ found by
%each seed are searched in the counterpart seed complete result.

%An overlap criterion on both part of each alignment is the criterion
%for non-exclusive alignments.

\end{document}